# Tunable Symmetries of Integer and Fractional Quantum Hall Phases in Multi-Dirac-Band Heterostructures

Petr Stepanov[1], Yafis Barlas[1], Tim Espiritu[1], Shi Che[1], Kenji Watanabe[2], Takashi Taniguchi[2], Dmitry Smirnov[3], Chun Ning Lau[1*]

[1] Department of Physics and Astronomy, University of California, Riverside, Riverside, CA 92521
[2] National Institute for Materials Science, 1-1 Namiki Tsukuba, Ibaraki 305-0044, Japan.
[3] National High Magnetic Field Laboratory, Tallahassee, FL 32310

**Abstract**
The co-presence of multiple Dirac bands in few-layer graphene leads to a rich phase diagram in the quantum Hall regime. Using transport measurements, we map the phase diagram of BN-encapsulated ABA-stacked trilayer graphene as a function charge density $n$, magnetic field $B$ and interlayer displacement field $D$, and observe transitions among states with different spin, valley, orbital and parity polarizations. Such rich pattern arises from crossings between Landau levels from different subbands, which reflect the evolving symmetries that are tunable *in situ*. At $D$=0, we observe fractional quantum Hall (FQH) states at filling factors 2/3 and -11/3. Unlike those in bilayer graphene, these FQH states are destabilized by a small interlayer potential that hybridizes the different Dirac bands.

Graphene[1-5] and its few-layer counterparts[3, 4, 6, 7] have emerged as fascinating platforms for investigating quantum Hall (QH) physics. In addition to their high mobility, the presence of spin and valley degrees of freedom leads to SU(4) symmetries that do not have counterparts in traditional two-dimensional electron gas based on GaAs heterostructures. In particular, in few-layer graphene, the addition of layers and multiple bands with non-trivial Berry curvatures provide additional degrees of freedom, resulting in an incredibly rich display of symmetries that can be broken either spontaneously or by external scalar or gauge fields.

One example of such a system is ABA-stacked trilayer graphene (TLG)[7-20]. In the absence of an interlayer displacement field $D$, the mirror reflection symmetry of the TLG allows its band structure to be decomposed into the massless mono-layer-graphene(MLG)-like and massive bilayer-graphene(BLG)-like fermions[22], which exhibit odd and even parities under mirror reflection, respectively. The presence of $D$ significantly impacts the band structure – it breaks the mirror reflection symmetry of the lattice, leading to hybridization of the MLG- and BLG-like bands; for sufficiently large $D$, new Dirac points are expected to emerge[24]. Thus $D$ is an extremely important parameter that allows *in situ* modification of TLG band structure; in the QH regime, it may modify the strengths of interactions and symmetries of Landau levels (LLs). However, unlike MLG and BLG, experimental studies of ABA-stacked TLG have been relatively limited[25-31], and most prior experiments are performed on singly-gated devices without proper control of layer polarization. Our prior work on dual-gated suspended devices allowed control of $D$[28], though the range of applied $D$ was rather limited due to the fragility of free-standing devices.

Here we study the magneto-transport properties of dual-gated ABA-stacked trilayer graphene (TLG) sandwiched between hexagonal BN (hBN) sheets, with mobility values up to 100,000 cm$^2$/Vs. In the QH regime, we observe a series of "ring"-like structures in the $R_{xx}(n, B)$ phase diagram, which evolve as $D$ increases (here $R_{xx}$ is the longitudinal resistance, $n$ the charge density and $B$ the magnetic field). Using a continuum $k \cdot p$ Hamiltonian that includes all remote-hopping parameters[24, 32], we determine that these ring-like patterns arise from crossings of between the $N$=0 MLG-like and $N \geq 2$ BLG-like LLs (here $N$ is the four-fold degenerate LL index); as $n$, $B$ and $D$ are modulated, the integer QH states transition between different spin, valley, orbital and parity (MLG- or BLG-like) symmetries[33]. Furthermore, we observe fractional QH (FQH) states at $\nu$=-11/3 and 2/3, which are attributed to parent integer states in BLG-like blocks. Intriguingly, unlike those in BLG that are only resolved at finite $D$[34, 35], these FQH states are only resolved at $D$=0, and destabilized by a very small interlayer potential, suggesting that they are extremely sensitive to the hybridization of the even- and odd-parity blocks. Our results highlight that various symmetries of QH states can be experimentally tuned *in situ* in ABA TLG, and shed light on the interplay between the competing degrees of freedom associated with layer, valley, spin and parity.

---

[*] Email: lau@physics.ucr.edu

hBN-encapsulated TLG devices are fabricated using techniques described in reference[36]. Briefly, TLG sheets are obtained via mechanical exfoliated and identified by color contrast in optical microscope and Raman spectroscopy[31, 37, 38]. hBN/TLG/hBN stacks are assembled using a dry pick-up technique[36], etched into Hall bar geometry, and coupled to Cr/Au electrodes (Fig. 1a). During the fabrication of our high-quality stacks we intentionally misalign the TLG and hBN sheets to avoid formation of Moire superlattices, which is also verified by the absence of secondary Dirac points (see Fig. 1b and 1c). A local graphite gate is placed underneath the conduction channel, so that the Fermi levels of the channel and the leads can be separately tuned by the graphite back gate and the Si gate, respectively[34]. The devices are measured in a He$^3$ cryostat at base temperature of 260mK.

Fig. 1b plots the four-terminal device resistance $R$ at $B$=0 as a function of charge density $n$ and out-of-plane displacement field $D$, where $n = (C_{bg}V_{bg} + C_{tg}V_{tg})/e$ and $D = \frac{(C_{bg}V_{bg} - C_{tg}V_{tg})}{2\varepsilon_0}$. Here $C_{bg}$ ($C_{tg}$) is capacitance per unit area between the TLG and the back (top) gate, $V_{bg}$ ($V_{tg}$) is the voltage applied to the back (top) gate, $e$ the electron charge, and $\varepsilon_0$ the permittivity of vacuum. We note that $D$ is the displacement field as imposed by the external gates, and, due to screening, is expected to be much larger than the actual electric field between the top and bottom layers. At $D$=0, $R$ is sharply peaked at $n$=0, and decreases precipitously to zero and even negative for large $n$ (Fig. 1c) indicating ballistic transport[5, 25] At $n$=0, $R(D)$ increases symmetrically with $D$ (Fig. 1d), a signature of low disorder in the device[10].

At high magnetic fields, the cyclotron orbits of the charges coalesce to form LLs. Fig. 2a displays the Landau fan $R_{xx}(n, B)$ at $D$=0. At $B$=8T, all integer QH states -18<$\nu$<18 as well as four-fold degenerate states up to $\nu$=±42 are resolved, indicating the high device quality. The QH states appear as dark blue bands with $R_{xx}$~0 that fan out from $(n,B)$=(0,0). Strikingly, a group of "ring"-like pattern appears, centering at $n$~5x10$^{11}$ cm$^{-2}$ and $B$~5T, as indicated by the dotted oval in Fig. 2a. Similar patterns are observed in all 5 ABA TLG samples studied to date. Interestingly, the QH states surrounding these "rings" exhibit a number of intriguing anomalies. For instance, the integers states between $\nu$=2 and 6 are not resolved for $B$<5T, *i.e.* before the appearances of the "rings"; however, after the emergence of the "rings" or $B$>5.5T, they are clearly resolved. On the other hand, the QH states between $\nu$=6 and 10 display the opposite behavior – they are clearly resolved before the rings, and unresolved for $B$>5.5T. Further examination of the data reveals a series of similar (albeit less resolved) features at $B\sim 3$T for states $10<\nu<14$, at $B\sim 2$T for $14<\nu<18$, at $B\sim 1.44$T for $18<\nu<22$, etc, with alternating resolution and disappearance of the associated QH states. Such disappearing and re-entrant QH states indicate abrupt changes in the LL gaps, hence suggesting that, despite at constant filling factors, these states undergo transitions between LLs with different symmetries[39, 40].

To quantitatively account for the data, we calculate TLG's LL spectrum at $D=0$ using a $k\cdot p$ continuum Hamiltonian to calculate the LL spectrum[24, 32], assuming spinless particles (Fig. 2b). Values of the hopping parameters are carefully chosen so as to match the LL crossing points, and are as follows: $\gamma_0$=3.23 eV, $\gamma_1$=0.31 eV, $\gamma_2$=-0.032 eV, $\gamma_3$=0.3 eV, $\gamma_4$=0.04 eV, $\gamma_5$=0.01 eV, $\delta$=0.027 eV, and $\delta_2$=1.8 meV. Here the $\gamma$'s are the Slonczewski– Weiss–McClure parameters of graphite[41], $\delta$ is the energy difference between stacked and non-stacked atoms, and $\delta_2$ is the potential difference between the middle layer and the average of the outer layers. The LLs are labeled by (M/B, $N$, ±), where M and B represent MLG- and BLG-like blocks, $N$ is the LL index, and +/- refers to the K/K' valleys, respectively. The LL spectrum can be decomposed into the MLG-like and BLG-like branches, represented by orange and blue lines, respectively. Unlike MLG and BLG, the valley degeneracy of TLG is not protected by inversion symmetry, and remote hopping terms lead to valley-splitting of the LLs. In particular, the $N$=0 LL of the MLG-like branch is valley-split into two non-dispersive states, with energy of the K valley LL ~2 meV higher than that of the K' valley; these LLs cross the BLG-like LLs at energies $E=\delta - \frac{\gamma_5}{2}$ and $-\gamma_2/2$, respectively. For the BLG-like block, the $N$=0 and $N$=-1 LLs of the BLG-like block are weakly dispersive, with valley splitting that is proportional to $B$. For higher LLs, the splitting is relatively small with the K-valleys occupying highest energies.

To better compare the data with theoretical calculations, we calculate the total density of states (DOS) based on LL spectrum shown in Fig. 2b, assuming each level is broadened by $\Gamma$=0.3meV, and Zeeman-split into up (↑) and down (↓) spins, with energy difference $g\mu_B B$, where $\mu_B$ is the Bohr magneton, and $g\sim 2$ is the $g$-factor. The resultant simulation is shown in Fig. 2c, and can satisfactorily account for all the major features. For instance, the "rings" around $B\sim 5$T arise from crossings between (M,0,±) and (B,2,±) LLs, and reflect the changing symmetries of the QH states. Let us first examine the QH states at $B$ immediately before the emergence of the "rings". Here the highest filled LLs for QH states $\nu$=3-6 are (B, 2, ±, ↑/↓); the energy gaps between these LLs are rather small, $<\sim 0.4$ meV, thus these QH states are barely resolved at $B\sim 4.5$T. As $B$ increases, the MLG-like and BLG-like branches cross, giving rise to the rings. As $B$ increases further past the "rings", the highest filled LLs transition to (M, 0, ±, ↑/↓). These MLG-like LLs have larger energetic separations, thus the QH states at $2<\nu<6$ become resolved. On the other hand, the QH states between $\nu$=6 and 10 undergo the opposite transition: as $B$ increases, the highest filled LLs for these state transitions from (M, 0, ±, ↑/↓) at $B\sim 4.5$T to (B, 2, ±, ↑/↓) at $B>5.5$T; similar to the states at lower filling factors, the different LL gaps within the (M, 0) and (B,2) blocks account for the resolution and disappearance of these QH plateaus.

Taken together, the co-presence of two different Dirac bands gives rises to a rich phase diagram, in which the various symmetries associated with QH states (parity, orbital, spin and valley) can be modified by *in situ* tuning $n$ and $B$. A map of these evolving symmetries of the observed QH states with $n$ and $B$ is sketched in Fig. 2d.

In addition to $n$ and $B$, the LL spectrum and hence the symmetries of QH states can also be tuned by $D$. To this end, we measure $R_{xx}(n,B)$ at different applied $D$. Two representative plots at $D$=84 meV/nm and 180 meV/nm are shown in Fig. 3a-b. Comparing to the data at $D$=0, one salient difference is that the LL crossing points first "contract" vertically so that they occur over a rather narrow range in $n$ at $D$=84 meV/nm, but "expand" to cluster around two different $n$ values for the larger displacement field, as indicated by the red arrows. Furthermore, an additional crossing point emerge near $\nu$=12, as indicated by the black arrows.

To understand these features, we calculate the total DOS in the ($n,B$) plane under application of an interlayer potential $\Delta$. By considering the LL crossing points, we determine that the experimental data are best accounted for by using $\Delta$~8.5 mV and 20 mV, respectively. The simulation results are shown in Fig. 3c-d, in very good agreement with the data. We note that technically, $\Delta$ hybridizes the MLG- and BLG-like bands, thus parity is no longer a good quantum number. Nevertheless, we can still identify the evolution of bands under small perturbation of $\Delta$ and determine whether they originate from MLG- or BLG-like blocks. We also note that screening significantly reduces the effective interlayer potential from $Dd$, i.e. $\Delta$~$Dd/6$ at n~$10^{12}$ cm$^{-2}$, where $d$~0.67nm is the distance between the outmost layers. This screening effect is particularly important at high doping, as the linearly growing density of states of the MLG-like block becomes comparable to and even surpasses that of the BLG-block as $n$ increases.

The evolving LL crossing patterns with $D$ belie the changing symmetries of the QH states, which can be determined by considering the effect of $\Delta$ on the LL spectrum. In general, $\Delta$ tends to elevates the $N\geq 0$ LLs originating from the K' valley relative to those from K valley (see Supplemental Materials). Such effect is most dramatic for the (M,0) LLs: at $\Delta$=0, the K-valley LL is higher in energy by ~2 meV; as $\Delta$ increases, these 2 LLs moves closer together and becomes almost degenerate at $\Delta$~8 mV, leading to LL crossings occurring over a relatively narrow range in $n$, as seen in Fig.s 3a and 3c. For even larger $\Delta$, the 2 LLs moves apart again, with energy of the K'-valley LL surpassing that from the K valley by 7meV at $\Delta$=20 mV and $B$=7T, hence the crossing points move further apart in $n$, as observed in Fig. 3b and 3d. Similarly, for the $N$=3 BLG-like block, LLs are valley-split by ~0.5 meV at $\Delta$=0. $\Delta$ depresses the K-valley LL while elevating K'-valley, causing them to cross at $\Delta$~10 mV. Combined with Zeeman splitting that scales with $B$, this gives rise to a new crossing point near $\nu$=12 that corresponds to transition from a valley-polarized to spin-polarized state; this crossing point moves to high $B$ as $\Delta$ increases. This is observed experimentally – as indicated by the black arrows, the crossing point occur at $B$~3T in Fig. 3a and $B$~6T in Fig. 3b. The symmetries of the highest filled LLs for QH states 2<$\nu$<18 at different $D$ are sketched in Fig. 3e-f.

Lastly, we focus on FQH states in ABA-TLG. A number of recent works reported observation of FQH effect in MLG and BLG[29, 34, 42-45], yet FQHE in TLG has not been conclusively demonstrated[34], nor its dependence on $D$ studied. With the exception of ref. [46], theoretical studies of FQH states in TLG are virtually non-existent. Here we present evidence for *tunable* FQH states in TLG from two devices. The top panel of Fig. 4a plots $\Delta R_{xx}(\nu, D)$ at constant $B$=7T for device 1 with high hole mobility (~50,000 cm$^2$/Vs), where $\Delta R_{xx}$ is the longitudinal resistance from which a smooth background is subtracted. Strikingly, close to $D$=0, a "half loop"-like feature appears within the $\nu$=-4 integer plateau. At $D$=0, $\Delta R_{xx}(\nu)$ exhibits two dips – a large dip, labeled "A", appearing within the "loop" at $\nu$=-4.1, and a smaller dip "B" outside the loop at $\nu$=-3.7 (Fig. 4a middle panel). This loop is completely absent for $D$>16 mV/nm, as evidenced from the line trace $\Delta R_{xx}(\nu)$ at $D$=30 mV/nm (bottom panel). As we determine that screening could reduce the effect of displacement field by a factor of ~6, the critical interlayer potential that destroys these features is indeed very small, ~ 1.8 mV. Both features A and B are also visible in $R_{xx}(n, B)$ data at $D=0$, but vanish when similar data are taken with top gate grounded where $D$ is uncontrolled (Fig. 4b).

The presence of these features at constant filling factors excludes localization or mesoscopic fluctuation-induced effects, but rather points to the formation of LLs. As $D$ increases, the minimum of feature $B$ disappears adiabatically, whereas feature A displays peaks in $R_{xx}$ at finite $D\sim\pm15$ mV/nm, suggesting transitions between states with different symmetries, similar to those observed for $\nu=2$ and $\nu=1$ states in BLG[35, 47, 48].

Since features A and B appear within the $\nu=-4$ plateau, at least one of the two corresponds to a FQH state; however, we are unable to unequivocally determine the nature of these states from $R_{xy}$ due to the limited resolution. Nevertheless, as feature A exhibits a deeper minimum, hence suggesting a larger LL gap, we tentatively assign feature A to the integer $\nu=-4$ state and B to the FQH state at $\nu=-11/3$. The slight offset of the dips from the exact filling factor values may be accounted for by the close proximity of the two states that each has considerable LL broadening. Also, other more exotic states, such as bubbles or stripes phases[49], cannot be definitively excluded; further experimental and theoretical studies will be necessary to ascertain the nature(s) of these states.

Fig. 4c presents data from device 2 with good electron mobility ($\sim$10,000 cm$^2$/Vs), plotting $R_{xx}(n, B)$ at $D=0$ and 30 mV/nm, respectively. A fractional state at $\nu=+2/3$ is visible at $B>\sim$42T at $D=0$, but is absent at finite $D$[50].

Such $D$-*tunable* FQH states have been observed in BLG [34, 35], but with one important distinction –the FQH states in BLG are observed only when $D$ *exceeds* certain critical values, whereas that in TLG only when $D$ is *below* a very small critical value. In fact, these are the only known FQH states that are destabilized by $D$. From our calculations, we know that at $D=0$, the QH states between $\nu=-6$ and 2 originate from the $N=0$ BLG-like blocks. Thus we tentatively attribute both FQH states to the BLG-like bands. More specifically, the $\nu=+2/3$ QH state is spin-polarized and resides in the middle layer, whereas the $\nu=-11/3$ is also spin-polarized, layer coherent state of the outermost layers, both with the orbital index $n=0$. Their destruction by $D$ is consistent with the breaking of mirror reflection symmetry and hybridization between the MLG- and BLG-like LLs, which modifies the orbital wave functions of the states.

**Acknowledgement**

We thank Denis Bandurin, Jeil Jung and Klaus von Klitzing for helpful discussions. The experiments are supported by DOE BES Division under grant no. ER 46940-DE-SC0010597. Y.B. is supported by SHINES, which is an Energy Frontier Research Center funded by DOE BES under Award # SC0012670. Growth of hexagonal boron nitride crystals was supported by the Elemental Strategy Initiative conducted by the MEXT, Japan and a Grant-in-Aid for Scientific Research on Innovative Areas "Science of Atomic Layers" from JSPS. Part of this work was performed at NHMFL that is supported by NSF/DMR-0654118, the State of Florida, and DOE.

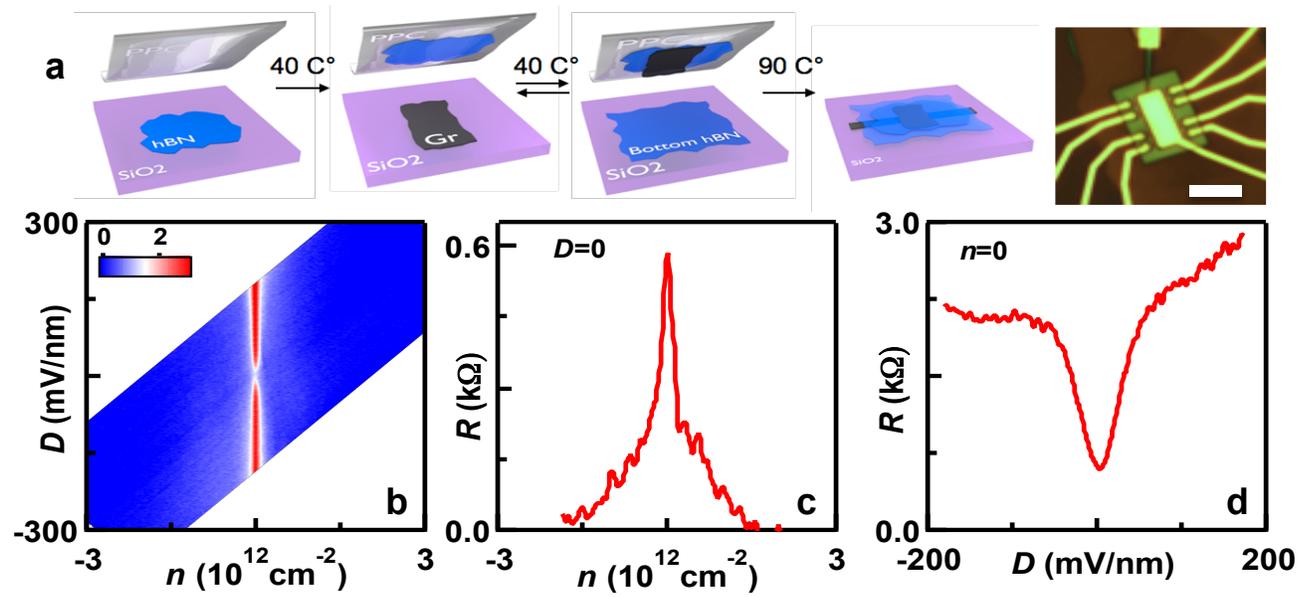

**Fig. 1. (a).** Schematics of fabrication and optical image of a finished device. Scale bar: 5 μm. **(b).** $R(n,D)$ in kΩ of an ABA TLG device at $B=0$. **(c-d).** Line traces $R(n)$ at $D=0$ and $R(D)$ at $n=0$.

**Fig. 2.** (a). $R_{xx}(n, B)$ in kΩ phase diagram at $D=0$. Numbers indicate filling factors. (b). Calculated LL spectrum using parameters listed in main text. (c). Simulated DOS$(n,B)$. (d). Schematics of symmetries of the highest filled LLs for QH states. The dark and light blue (orange) bands indicate K and K' valleys of the BLG-like (MLG-like) block, white bands indicate experimentally unresolved QH states, and arrows indicate spins. Inset: Zoomed-in schematics of the dotted square in (b), showing the crossings between the (M,0,±) and (B,2,±) LLs with Zeeman splitting.

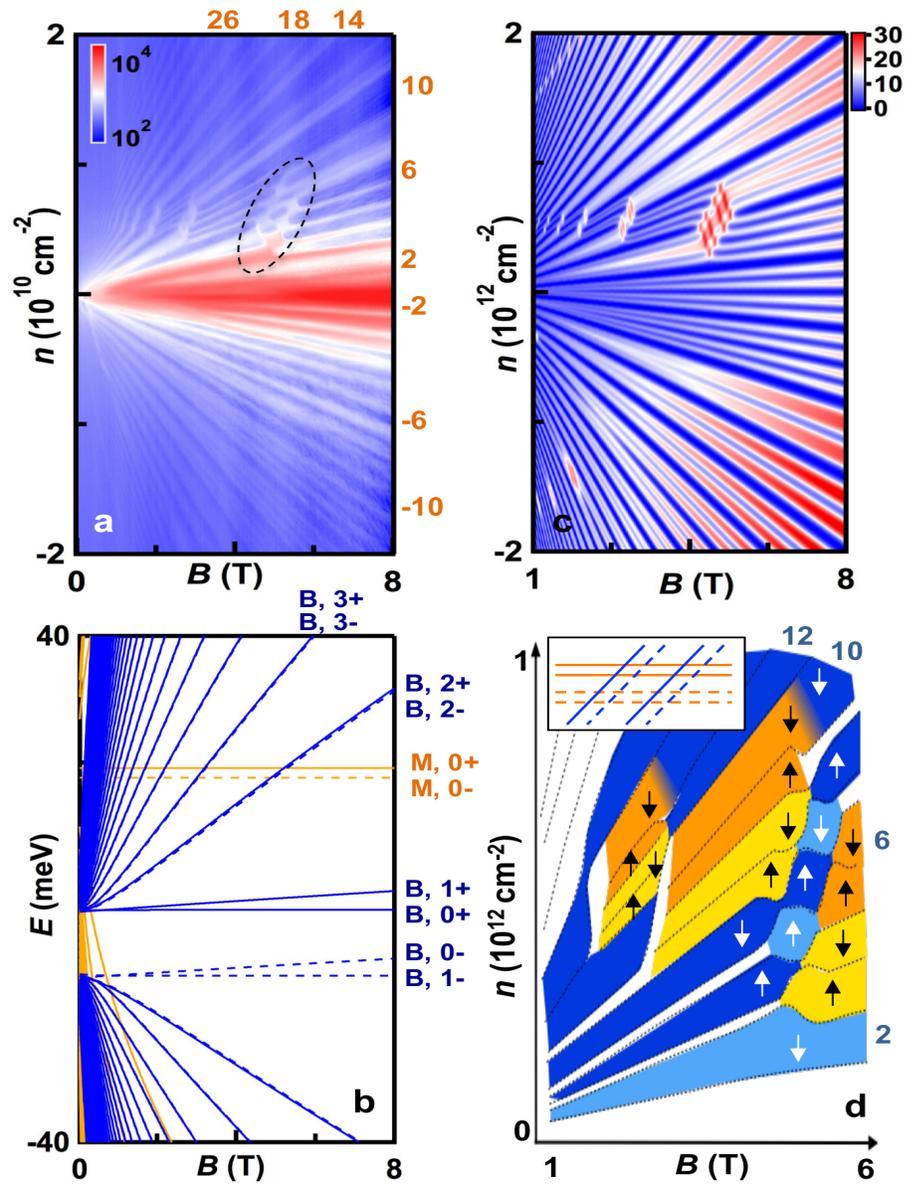

**Fig. 3.** (a-b). $R_{xx}(n,B)$ in kΩ at $D$=84 and 180 mV/nm, respectively. (c-d). Simulated DOS$(n,B)$ at $\Delta$=8 mV and 20 mV. (e-f). Schematics of symmetries of the highest filled LL for the QH states shown in (a-b). Color schemes are same as Fig. 2. The numbers indicate filling factors. Inset: Zoom-in schematics of the crossings between the (M,0,±) and (B,2,±) LLs.

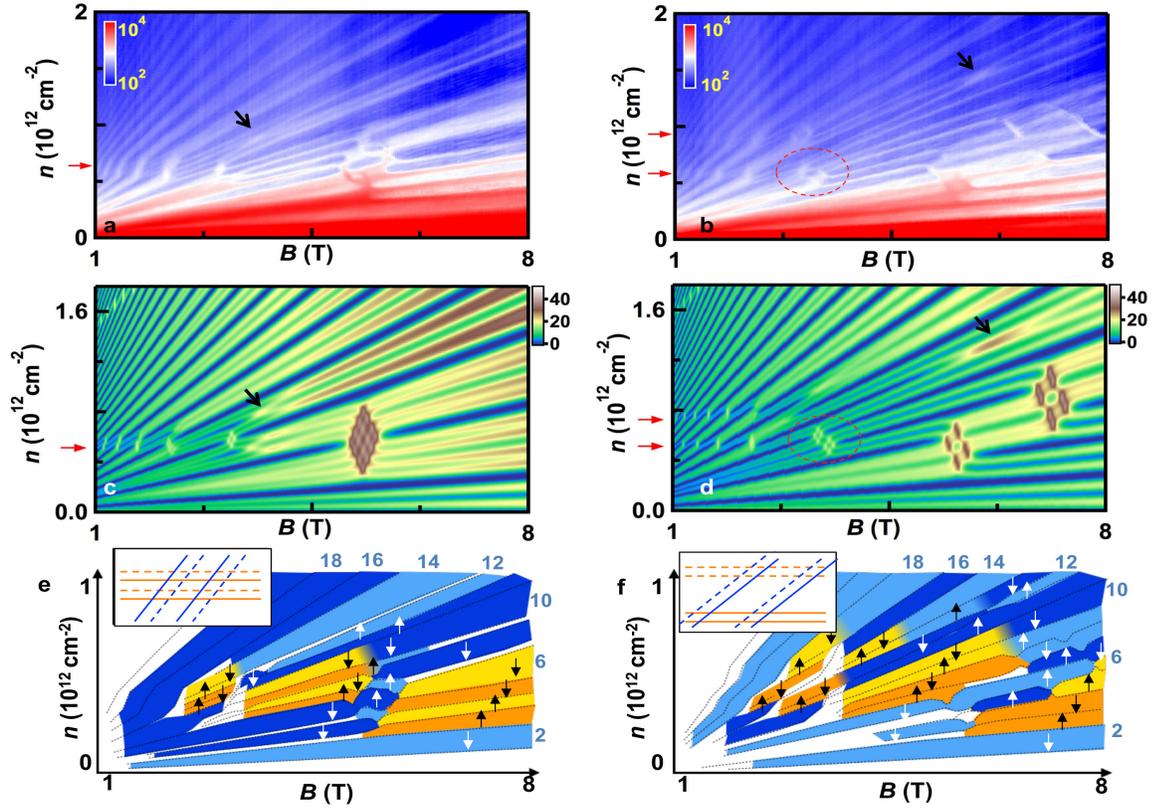

Fig. 4. Tunable fractional phases in TLG. (a). $\Delta R_{xx}(D, \nu)$ and line traces $\Delta R_{xx}(\nu)$ at $D=0$ and 30 mV/nm from Device 1 at $B=7$T. (b). $R_{xx}(n, B)$ from Device 1. Left panel is taken at $D=0$, and right panel is taken with top gate grounded ($D$ is uncontrolled). (c). $R_{xx}(n, B)$ from Device 2 at high magnetic field at $D=0$ (top panel) and 30 mV/nm (bottom). The $\nu=2/3$ FQH state is indicated by the red arrow. (All panels) Numbers indicate filling factors. Color scales are in units of k$\Omega$.

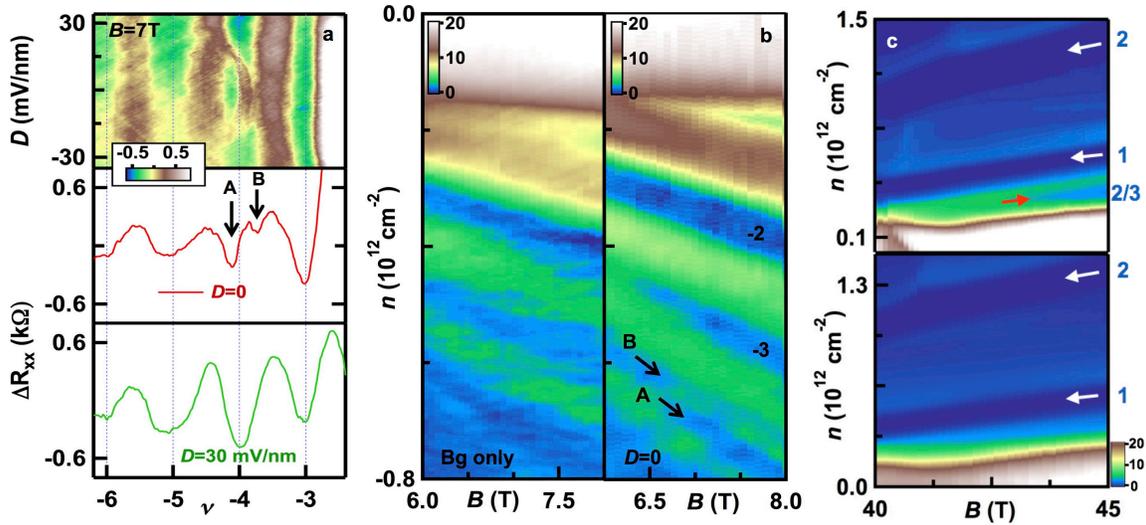